\documentclass[12pt]{article}
\usepackage[utf8]{inputenc}

\usepackage{authblk}

\title{Small Area Estimation with Random Forests and the LASSO}
\author[1]{Victoire Michal}
\author[2]{Jon Wakefield}
\author[1]{Alexandra M. Schmidt}
\author[3]{Alicia Cavanaugh}
\author[3]{Brian Robinson}
\author[1,4]{Jill Baumgartner}

\affil[1]{Department of Epidemiology, Biostatistics and Occupational Health, McGill University, Montreal, Canada}

\affil[2]{Department of Biostatistics, University of Washington, Seattle, USA}

\affil[3]{Department of Geography, McGill University, Montreal, Canada}

\affil[4]{Institute for Health and Social Policy, McGill University, Montreal, Canada}

\date{}

\usepackage[margin=1in]{geometry}
\usepackage{graphicx}
\usepackage{subfig}
\usepackage{float}
\usepackage{bm}
\usepackage{xurl}
\usepackage{amsmath,amsthm, amsfonts, dsfont}
\usepackage{multirow}
\usepackage{booktabs}
\usepackage{natbib}
\usepackage{color, xcolor}
\definecolor{gray}{rgb}{0.5,0.5,0.5}

\usepackage[shortlabels]{enumitem}

\usepackage{setspace}
\usepackage{listings}
\lstdefinelanguage{R}{
morekeywords={},
}
\begin{document}
%\fontsize{12pt}{20pt}\selectfont

% For JSSM: comment \maketitle and leave title centered and vspace
\maketitle

%\begin{center}
%    {\LARGE Small Area Estimation with Random Forests and the LASSO
%    }
%\end{center}

%\vspace{0.7cm}

\begin{abstract}
We consider random forests and LASSO methods for model-based small area estimation when the number of areas with sampled data is a small fraction of the total areas for which estimates are required. Abundant auxiliary information is available for the sampled areas, from the survey, and for all areas, from an exterior source, and the goal is to use auxiliary variables to predict the outcome of interest. We compare areal-level random forests and LASSO approaches to a frequentist forward variable selection approach and a Bayesian shrinkage method.
Further, to measure the uncertainty of estimates obtained from random forests and the LASSO, we propose a modification of the split conformal procedure that relaxes the assumption of identically distributed data. 
This work is motivated by Ghanaian data available from the sixth Living Standard Survey (GLSS) and the 2010 Population and Housing Census. We estimate the areal mean household log consumption using both datasets. The outcome variable is measured only in the GLSS for 3\% of all the areas (136 out of 5019) and more than 170 potential covariates are available from both datasets. Among the four modelling methods considered, the Bayesian shrinkage performed the best in terms of bias, MSE and prediction interval coverages and scores, as assessed through a cross-validation study. We find substantial between-area variation, the log consumption areal point estimates showing a 1.3-fold variation across the GAMA region. The western areas are the poorest while the Accra Metropolitan Area district gathers the richest areas.
\end{abstract}


\vspace{0.7cm}

\section{Motivation}
\label{sec:Motivation}

%%%%%%%%%%%%%%%%%%%%%%%%%%%%%%%
%
% INTRO
%
%%%%%%%%%%%%%%%%%%%%%%%%%%%%%%%

In 2015, the United Nations (UN) released their 2030 agenda for sustainable development goals (SDGs) consisting of 17 goals, the first of which was to end poverty worldwide \citep{resolution2015transforming}. %\textcolor{red}{Extreme poverty is measured as an individual's income lower than 1.25\$ a day \citep{SDG1}.} 
For their first SDG, the UN made seven guidelines explicit, including the implementation of ``poverty eradication policies" at a disaggregated level. To that end, producing reliable and fine-grained pictures of socioeconomic status and income inequality is fundamental to help decision makers prioritise and target certain areas. These detailed maps help local communities understand their situation compared to their neighbours, which also helps when planning interventions \citep{bedi2007more}.  

In Ghana, household surveys are collected every few years to measure the living conditions of households across Ghanaian regions and districts and to monitor poverty. To keep track of the Ghanaian population wealth, the equivalised consumption is recorded for the sampled households. Although the household income is not directly measured, the equivalised expenditure is an alternative that allows decision makers to assess a household's standard of living \citep{johnson2005economic}. This measure corresponds to the household consumption scaled by a weight based on the number of members in the household. We aim to estimate the equivalised consumption at a disaggregated level, to help policymakers better understand the distribution of the households' living standard in Ghana, in order to prioritise certain areas when implementing poverty eradication policies. The sixth Ghana Living Standards Survey (GLSS), conducted in 2012-2013, was the last household survey carried out prior to the new UN SDGs agenda. The fifth GLSS had shown that inequalities had increased since 2006. In particular, although the overall poverty decreased nation-wide, the wealthiest decile of the population consumed 6.8 times more than the poorest \citep{cooke2016ghana}. A downside of these household surveys is that the sampling design is stratified two-stage, which only allows for reliable survey sampling estimates at the district level, at best. Ghana is divided into 10 regions, which are formed by 170 districts or, at a finer level, around 38,000 enumeration areas (EAs). Producing reliable estimates at the EA level would further help the authorities in their policy decisions. 

We analyse data from the sixth GLSS for the Greater Accra Metropolitan Area (GAMA), which consists of 8 districts. The GLSS used a stratified two-stage cluster sample in which strata are defined by an urban or rural indicator. Then, the clusters, which correspond to the EAs, were sampled following proportional to size sampling. Within the sampled EAs, 15 households were systematically sampled. For each sampled household, we have detailed assessment of consumption and their level of education, employment, assets, with, in total, 174 auxiliary variables. This gives a sample of 136 EAs out of the 5019, in this Ghanaian region. The sampled EAs are anonymised, which means it is unknown which 136 EAs of the 5019 EAs are represented in the survey. Additionally, we have data available from the 2010 Ghanaian census for all EAs in the GAMA. Among others, the same 174 variables are measured in this census and in the sixth GLSS. The aim of this work is to produce estimates with uncertainty of the mean log household consumption at the EA level in the GAMA. 

In this paper, to deal with the higher number of auxiliary variables compared to the number of sampled EAs, we assess the performance of random forests and the LASSO (which performs variable selection) to estimate the mean household log consumption at the EA level in the GAMA. For the sake of comparison, we also consider a forward variable selection approach in the frequentist framework and a Bayesian shrinkage method. For all four approaches, we adopt EA-level models. Further, we propose a modification of the split conformal procedure to compute prediction intervals for the random forest and LASSO predictions while relaxing the assumption of identically distributed responses, which is necessary due to the complex sampling design. 

This paper is organised as follows. Section \ref{sec:LitReview} briefly reviews the literature on small area estimation (SAE) and variable selection in the frequentist, Bayesian and machine learning frameworks. Section \ref{sec:Model} describes the four methods that will be compared and the proposed procedure to produce prediction intervals for estimates obtained through random forests and the LASSO. Section \ref{sec:Analyses} shows the results from two simulation studies. First, Section \ref{sec:Sim_scaledSC} presents a comparison between the proposed modified split conformal and the original split conformal procedures. Then, Section \ref{sec:Sim_comparison} provides a comparison between the four methods that perform variable selection. Section \ref{sec:GAMA} discusses the results from the four methods applied to the Ghanaian datasets. Finally, Section \ref{sec:Discussion} concludes the paper with a discussion.

\subsection{Literature review}
\label{sec:LitReview}

SAE concerns estimation of area-level summaries when data is sparse or non-existent in the areas \citep{rao2015small}. This area of research in survey sampling has greatly evolved in the last 50 years \citep{pfeffermann2002small, pfeffermann2013new, rao2015small, ghosh2020small}. \cite{tzavidis2018start} points out that the use of SAE by national statistical institutes (NSIs) and other organisations to produce official statistics exhibits this increasing popularity; e.g., the \texttt{povmap} software developed by the World Bank \citep{elbers2003micro, bank2015software} and the Small Area Income and Poverty Estimates project carried out by the US Census Bureau \citep{bureau2018small}. 

In survey sampling, the design-based framework may be distinguished from the model-based framework. Design-based methods, also called randomisation methods, assume the variable of interest to be fixed in the finite population while the randomness comes from the sampling process. 
Direct (weighted) estimators have favourable design-based properties in large samples and rely only on the sampling weights and the recorded responses within each sampled area to produce areal estimates. Hence, estimates for non-sampled areas are missing. Additionally, data sparsity yields imprecise direct estimates at the areal level. Similarly, data sparsity within areas may lead to imprecise model-assisted estimates. These latter approaches also fall under the umbrella of design-based inference. Model-assisted methods are design-based approaches which model the responses to gain precision but are still design consistent \citep{sarndal2003model}. An alternative is to use model-based approaches. In this setting, the responses are no longer assumed fixed but treated as random variables which are modelled using auxiliary information and/or random effects. In model-based methods for SAE, it is common to use exterior sources of information to augment the auxiliary information from the sample to the entire finite population; for example, using information obtained from censuses. \cite{tzavidis2018start} describe a two-step approach to produce model-based small area estimates. First, a model is fitted using the survey responses and survey auxiliary variables. Then, the outcome is predicted for the entire finite population according to the estimated model parameters and finite population auxiliary information. 
%Further, modelling approaches can be categorised into unit-level methods and area-level methods \citep{rao2015small}. In areal-level models, the outcome and covariates are aggregated to form direct estimates at the areal level before modelling. In unit-level models, unit-level responses are linked to unit-level auxiliary variables. However, it may be difficult to obtain such information for the entire finite population.  

Abundant auxiliary information may be measured in the sample, for the sampled areas, and through exterior sources, for all the areas of the region of interest. It may therefore be necessary to select a subset of covariates to model the response variable, in the presence of high-dimensional auxiliary information. In this way, precision can be increased as unnecessary auxiliary variables are not included. The inference procedure for model-based approaches can be performed under the frequentist or Bayesian frameworks, or with flexible parametric models via machine learning techniques.

Machine learning methods are becoming more and more popular in the survey sampling community; see for example, \cite{wang2014bagging} and \cite{breidt2017model}. However, it is not straightforward to perform inference and assess the estimates' uncertainty under these approaches. For example, the bootstrap does not work for non-smooth targets \citep{dezeure2015high}. Among machine learning methods, random forests \citep{breiman2001random} can be fitted to unit-level or area-level data for a flexible approach. Random forests are a collection of regression trees that recursively partition the responses into increasingly homogeneous subgroups (nodes), based on covariate splits. Random forests, which present the benefit of accommodating non-linear relationships and complex interactions, naturally select variables through these covariate splits. Each individual regression tree is fitted on a bootstrap sample of the original dataset. There is a growing literature on methods to measure the uncertainty of random forest point estimates. For instance, different Jackknife approaches have been proposed \citep{steinberger2016leave, wager2014confidence, wager2018estimation}. However, these procedures present drawbacks, such as their computational overheads. Additionally, it is unclear how they apply to survey data. Recently, \cite{zhang2019random} proposed the so-called out-of-bag (OOB) prediction intervals, which are computed based on quantiles of the random forest out-of-bag prediction errors. These denote the difference between a data point's outcome and its point estimate, obtained from a random forest grown without said data point. In simulation studies, \cite{zhang2019random} show that their proposed method performs similarly to the split conformal (SC) approach proposed by \cite{lei2018distribution}. The SC approach may be used to compute prediction intervals for any modelling method (e.g., linear models or random forests). To compute prediction intervals for random forest estimates through the SC method, the original dataset is first split into two datasets. A random forest is trained on one subsample, and point estimates and their associated prediction errors are obtained for the other subsample. Then, the intervals are computed based on the empirical quantiles of the prediction errors from the second subsample. Note that while the OOB method proposed by \cite{zhang2019random} only estimates prediction intervals for random forests, the SC method can be applied to any modelling procedure used to obtain point estimates. A common feature of all these prediction interval methods is that the data are assumed to be independently and identically distributed. This is a strong assumption and is not usually true for data gathered from a complex survey design.

Inference procedures for model-based approaches can also follow the frequentist or the Bayesian paradigms. In these frameworks, variable selection is an important yet contentious research topic. In the frequentist framework, two-step procedures are common. Variables are first iteratively selected (forward selection) or removed (backward elimination) to model the outcome, based on the optimisation of some criterion (e.g., AIC, BIC, $R^2$). Then, a final model that includes only the selected covariates is fitted to the data. In SAE, it is common to select variables by comparing models through some criterion (for example, AIC or BIC, or survey sampling adjusted versions); see e.g., \cite{han2013conditional, rao2015small} and \cite{lahiri2015variable}. In the frequentist framework, regularisation methods have also been proposed in the literature, such as ridge regression and the LASSO \citep{tibshirani1996regression, tibshirani2011regression, mcconville2017model}. These methods apply constraints to the regression parameters. However, in the case of the LASSO, these constraints yield estimates of the model parameters whose uncertainty estimation is difficult, especially in a survey setting. In a simulation study, \cite{lei2018distribution} show that their proposed SC method performs well in computing prediction intervals for predictions obtained through the LASSO, when the data are independently and identically distributed. 

In the Bayesian framework, variable selection is conducted by imposing informative priors on the model parameters. Multiple shrinkage priors have been proposed in the literature, for example, Bayesian ridge regression and the Bayesian LASSO \citep{hans2010model}. In the former, a Gaussian prior is assigned to the regression parameters, while a double-exponential distribution is used for the latter. It can be shown that, under the respective priors, computing the maxima \textit{a posteriori} to estimate the parameters results exactly in ridge-type and LASSO-type estimators \citep{reich2019bayesian}. A more recent popular approach \citep{carvalho2010horseshoe} is the use of the horseshoe prior, which imposes \textit{a priori} a heavier weight towards 0 than a normal or double-exponential distribution. 

\section{Methods}
\label{sec:Model}

%%%%%%%%%%%%%%%%%%%%%%%%%%%%%%%
%
% METHODS
%
%%%%%%%%%%%%%%%%%%%%%%%%%%%%%%%

Let a region be divided into $M$ non-overlapping areas, $A_c, \ c=1, \dots, M$. Denote by $N_c$ the number of units in $A_c$, with outcomes $y_{ck}, \ k=1, \dots, N_c$. The main goal is to estimate the areal mean $\overline{y}_c = (1/N_c)\sum_{k=1}^{N_c} y_{ck}$ for all areas $c=1, \dots, M$, using a sample of $n_c$ units taken from $c=1, \dots, m$ areas. Denote by $s$ the set of area and household indices included in the sample and denote by $s_c, \ c=1, \dots, M$, the set of sampled units in the $c$-th area. Let $f_c = n_c/N_c$ be the sampling fraction within each area. For any variable $a$, let $\overline{a}_c = (1/N_c)\sum_{k=1}^{N_c}a_{ck}, \ c=1, \dots, M$, be the population areal mean, and $\overline{a}_c^{(s)} = (1/n_c)\sum_{k \in s_c}a_{ck}, \ c=1, \dots, m$ and $\overline{a}_c^{(ns)} = (1/(N_c-n_c))\sum_{k \notin s_c}a_{ck} = \left(\overline{a}_c-f_c\overline{a}_c^{(s)}\right)/(1-f_c), \ c=1, \dots, M$, the areal means for the sampled (subscript $(s)$) and non-sampled (subscript $(ns)$) units, respectively. For all $M$ areas, the estimation target may be decomposed as follows 
\begin{equation}
    \overline{y}_c = \frac{1}{N_c} \left(\sum_{k \in s_c}y_{ck} + \sum_{k \notin s_c} y_{ck}\right) = f_c \overline{y}_c^{(s)} + (1-f_c)\overline{y}_c^{(ns)}, \ c=1, \dots, M.
\end{equation}
To estimate $\overline{y}_c$, the non-sampled mean, $\overline{y}_c^{(ns)}$, remains to be estimated for all $M$ areas. Let $\widehat{\overline{Y}}_c^{(ns)}, \ c=1, \dots, M$, be the estimator of $\overline{y}_c^{(ns)}, \ c=1, \dots, M$. The prediction approach estimator \citep{lohr2021sampling} for the target of inference is 
\begin{equation}
    \widehat{\overline{Y}}_c = f_c \overline{y}_c^{(s)} + (1-f_c)\widehat{\overline{Y}}_c^{(ns)}, \ c=1, \dots, M.
    \label{eq:estimator_general}
\end{equation}
The uncertainty of $\widehat{\overline{Y}}_c$ may be measured using prediction intervals of level $(1-\alpha)\%$, $\mathrm{PI}_{(1-\alpha)}$, of the form 
\begin{equation}
    \mathrm{PI}_{(1-\alpha)\%}\left[\widehat{\overline{Y}}_c\right] = f_c \overline{y}_c^{(s)} + (1-f_c) \mathrm{PI}_{(1-\alpha)\%}\left[\widehat{\overline{Y}}_c^{(ns)}\right], \ c=1, \dots, M.
    \label{eq:uncertainty_general}
\end{equation}
Note that for a non-sampled area $c'$, $f_{c'} =0$ and the estimator reduces to $\widehat{\overline{Y}}_{c'} = \widehat{\overline{Y}}_{c'}^{(ns)}$, with prediction interval, $\mathrm{PI}_{(1-\alpha)\%}\left[\widehat{\overline{Y}}_{c'}\right] = \mathrm{PI}_{(1-\alpha)\%}\left[\widehat{\overline{Y}}_{c'}^{(ns)}\right]$. 

Random forests and the LASSO are considered to estimate $\widehat{\overline{Y}}_c^{(ns)}$ in the model-based framework. For the sake of comparison, we also consider a forward variable selection approach in the frequentist paradigm and a Bayesian shrinkage method. In this model-based framework, the finite population response values $\{y_{ck}, \ c=1, \dots, M, \ k=1, \dots, N_c\}$ are assumed to be a realisation of super population independent random variables $Y_{ck}$ that follow the model
\begin{equation}
    \mathbb{E}(Y_{ck})=f(\bm{x}_{ck}, \bm{\beta}) \quad \mbox{and} \quad \mathbb{V}(Y_{ck})=\sigma^2,   \label{eq:modelbased_unit}
\end{equation}
where $\bm{x}_{ck}$ is a $p$-dimensional vector of covariates. Consequently, it is assumed that the non-sampled units follow the same model as the sampled units. 

The four modelling approaches assume there are covariates available from the sample, $\{\bm{x}_{ck}, \ c,k \in s\}$, as well as areal covariate means, $\overline{\bm{x}}_c, \ c=1, \dots, M$, which are known for all the areas of the finite population. Such information may be obtained from a census. Inference is carried out at the areal level in all three methods. The super population model (\ref{eq:modelbased_unit}) implies the following sampled and non-sampled moments
\begin{align}
    \mathbb{E}\left(\overline{Y}_c^{(s)}\right) &= f(\overline{\bm{x}}_c^{(s)}, \bm{\beta}), \quad &\mathbb{V}\left(\overline{Y}_c^{(s)}\right) &= \sigma^2/n_c, \notag \\   
    \mathbb{E}\left(\overline{Y}_c^{(ns)}\right) &= f(\overline{\bm{x}}_c^{(ns)}, \bm{\beta}), \quad &\mathbb{V}\left(\overline{Y}_c^{(ns)}\right) &= \sigma^2/(N_c-n_c).
    \label{eq:modelbased_meanmoments}
\end{align}
Therefore, inference is conducted using $\left\{\left(\overline{y}_c^{(s)}, \overline{\bm{x}}_c^{(s)}\right), \ c=1, \dots, m\right\}$ and the non-sampled mean predictions, $\widehat{\overline{Y}}_c^{(ns)}, \ c=1, \dots, M$, are computed using the available covariates' non-sampled means, $\overline{\bm{x}}_c^{(ns)}, \ c=1, \dots, M$.\\

\subsection{Machine learning approach}
First, we consider a random forest prediction approach. This non-parametric method makes no further assumption to Model (\ref{eq:modelbased_unit}). The corresponding moments of the sampled and non-sampled outcome means are as in (\ref{eq:modelbased_meanmoments}). 
Following \cite{breiman2001random}, random forest point estimates are the average over $B$ point estimates obtained from training $B$ independent regression trees on $B$ bootstrap versions of the original sample. Each regression tree partitions the bootstrap response values based on splitting rules applied to covariates. 
A random forest algorithm is described in appendix \ref{App:RF}. To measure the uncertainty associated to the random forest predictions, we propose a modification of the SC prediction intervals of \cite{lei2018distribution} which relaxes the assumption of identically distributed sampled and non-sampled data points. The original SC procedure assumes $\overline{Y}_c^{(s)}$ and $\overline{Y}_c^{(ns)}$ to be independently and identically distributed (i.i.d.). However, as shown in (\ref{eq:modelbased_meanmoments}), $\overline{Y}_c^{(s)}$ and $\overline{Y}_c^{(ns)}$ are not identically distributed. Hence, in the proposed modified SC procedure, we assume the mean structures to be similar and allow the variances to be scaled differently, as is the case in (\ref{eq:modelbased_meanmoments}). Specifically, in this context of a complex sampling design, we assume the variance is independent of the sample strata. The unit-level variance, $\sigma^2$, is assumed fixed across the strata and the sampled and non-sampled areal-level variances only vary with the number of sampled and non-sampled units, $n_c$ and $N_c-n_c$, respectively.
We propose to scale the residuals computed in the original SC procedure before computing the empirical quantile necessary to the prediction intervals. Said quantile is then scaled when computing the prediction intervals. The proposed scaled SC procedure can be described through the following steps:
\begin{enumerate}
    \item Randomly split $\left\{\left(\overline{y}_c^{(s)}, \overline{\bm{x}}_c^{(s)}\right), \ c=1, \dots, m\right\}$ into two equal sized datasets. Denote by $S_1$ and $S_2$ the resulting two sets of area indices;
    \item Train a random forest on $\left\{\left(\overline{y}_c^{(s)}, \overline{\bm{x}}_c^{(s)}\right), \ c \in S_1\right\}$ and predict $\left\{\widehat{\overline{Y}}_c^{(S_2)}, \ c \in S_2\right\}$;
    \item Compute the scaled absolute residuals $R_c = \sqrt{n_c} \times \left|\overline{y}_c^{(s)}-\widehat{\overline{Y}}_c^{(S_2)}\right|, \ c \in S_2$;
    \item Find $d_\alpha$, the $k_\alpha$th smallest residual $R$, for $k_\alpha = \lceil (m/2+1)(1-\alpha)\rceil$;
    \item Let the prediction interval be $\mathrm{PI}_{(1-\alpha)\%}\left[\widehat{\overline{Y}}_c^{(ns)}\right] = \widehat{\overline{Y}}_c^{(ns)} \pm d_\alpha/\sqrt{N_c-n_c}, \ c=1, \dots, M$.
\end{enumerate}

\noindent Hence, with a random forest procedure, the estimator and its uncertainty (\ref{eq:estimator_general}) and (\ref{eq:uncertainty_general}) become
$$\widehat{\overline{Y}}_c = f_c \overline{y}_c^{(s)} + (1-f_c) \left(\sum_{c'=1}^m w_{c'}(\overline{\bm{x}}_c^{(ns)}) \overline{y}_{c'}^{(s)}\right), \quad \mathrm{PI}_{(1-\alpha)\%}\left[\widehat{\overline{Y}}_c\right] = \widehat{\overline{Y}}_c \pm (1-f_c) \frac{d_\alpha}{\sqrt{N_c-n_c}},$$
where the weights $w_{c'}(\cdot)$ result from the random forest procedure shown in appendix \ref{App:RF}.

\subsection{Frequentist approach: the LASSO} 
Second, we consider the LASSO to predict the areal non-sampled means while performing variable selection. 
%Model (\ref{eq:modelbased_unit}) is completed with a normal distribution assumption: $Y_{ck} \overset{i.i.d.}{\sim} \mathcal{N}(\bm{x}_{ck}^\top\bm{\beta}, \sigma^2), \ \forall c,k$. Hence, the sampled and non-sampled means independently follow $\mathcal{N}(\overline{\bm{x}}_{c}^{(s)\top}\bm{\beta}, \sigma^2/n_c)$ and $\mathcal{N}(\overline{\bm{x}}_{c}^{(ns)\top}\bm{\beta}, \sigma^2/(N_c-n_c))$, respectively. 
The LASSO estimates $\widehat{\bm{\beta}}^{\mbox{\tiny LASSO}}$ by solving $\underset{\bm{\beta} \in \mathbb{R}^p}{\min}\left\{\left\|\overline{\bm{y}}^{(s)}-\overline{\bm{x}}^{(s)}\bm{\beta}\right\|_2^2/(2m) + \lambda\left\|\bm{\beta}\right\|_1\right\},$ $\lambda \geq 0,$ where $\overline{\bm{y}}^{(s)} = \left[\overline{y}_1^{(s)}, \dots, \overline{y}_m^{(s)}\right]^\top$ and $\overline{\bm{x}}^{(s)} = \left[\overline{\bm{x}}^{(s)^\top}_1, \dots, \overline{\bm{x}}_m^{(s)^\top}\right]^\top$. Note that the shrinkage penalty parameter $\lambda$ is fixed after a 10-fold cross-validation, seeking the smallest test MSE. To measure the uncertainty associated with these LASSO predictions, the proposed scaled SC approach described above is applied to obtain $d_\alpha^{\mbox{\tiny LASSO}}$, for an interval of level $(1-\alpha)\%$. Note that in the second step of the proposed approach, the LASSO is fitted to the first subsample $S_1$, instead of a random forest. Therefore, the estimator and its uncertainty (\ref{eq:estimator_general}) and (\ref{eq:uncertainty_general}) become $$\widehat{\overline{Y}}_c = f_c \overline{y}_c^{(s)}+(1-f_c)\left[\overline{\bm{x}}_c^{(ns)^\top}\widehat{\bm{\beta}}^{\mbox{\tiny LASSO}}\right], \quad \mathrm{PI}_{(1-\alpha)\%}\left[\widehat{\overline{Y}}_c\right] = \widehat{\overline{Y}}_c \pm (1-f_c) \frac{d_\alpha^{\mbox{\tiny LASSO}}}{\sqrt{N_c-n_c}}.$$

\subsection{Frequentist approach: forward variable selection} 
As a comparison, we consider a frequentist method with the commonly used forward approach with AIC as a variable selection criterion. Model (\ref{eq:modelbased_unit}) is completed by assuming the errors are normally distributed. 
%Hence, the sampled and non-sampled means independently follow $\mathcal{N}(\overline{\bm{x}}_{c}^{(s)\top}\bm{\beta}, \sigma^2/n_c)$ and $\mathcal{N}(\overline{\bm{x}}_{c}^{(ns)\top}\bm{\beta}, \sigma^2/(N_c-n_c))$, respectively. 
To predict $\widehat{\overline{Y}}_c^{(ns)}$, the forward approach is a two-step procedure. First, a subset of $K$ covariates $\bm{z}$ is selected among the available $\bm{x}$'s. To that end, linear models are iteratively fitted, adding one covariate at a time based on the resulting AIC value. Then, using the selected covariates, a linear model is fitted: $\overline{y}_c^{(s)} \sim \mathcal{N}\left(\overline{\bm{z}}_c^{(s)^\top}\bm{\eta}, \sigma^2/n_c\right), \ c=1, \dots, m$, to estimate $\widehat{\bm{\eta}}$, $\widehat{\mathbb{V}}\left(\widehat{\bm{\eta}}\right)$ and $\widehat{\sigma}$. The steps required to run this forward approach are detailed in Appendix \ref{App:AIC}. The estimator and uncertainty (\ref{eq:estimator_general}) and (\ref{eq:uncertainty_general}) become 
\begin{align*}
    &\widehat{\overline{Y}}_c = f_c \overline{y}_c^{(s)} + (1-f_c) \left[\overline{\bm{z}}_c^{(ns)^\top}\widehat{\bm{\eta}}\right], \\
    &\mathrm{PI}_{(1-\alpha)\%}\left[\widehat{\overline{Y}}_c\right] = \widehat{\overline{Y}}_c \pm q_\alpha (1-f_c)\sqrt{\overline{\bm{z}}_c^{(ns)^\top}\widehat{\mathbb{V}}\left(\widehat{\bm{\eta}}\right)\overline{\bm{z}}_c^{(ns)} + \frac{\widehat{\sigma}^2}{N_c-n_c}},
\end{align*}
where $q_\alpha$ denotes the $\alpha$-level quantile from a $\mathcal{N}(0,1)$ distribution. Note that uncertainty in the covariates selected in not accounted for.\\

\subsection{Bayesian approach}
Finally, a Bayesian approach is considered, where all the available covariates, $\bm{x}$, are used in a single step to model the outcome while applying the horseshoe prior \citep{carvalho2010horseshoe} to the regression parameters. Similar to the forward approach, a normal distribution is further assumed for Model (\ref{eq:modelbased_unit}). The observed sampled means are modelled through $\overline{y}_c^{(s)} \sim \mathcal{N}(\overline{\bm{x}}_{c}^{(s)\top}\bm{\beta}, \sigma^2/n_c), \ c=1, \dots, m,$ with priors $\beta_j \sim \mathcal{N}(0,\lambda_j^2 \tau^2), \ j=1, \dots, p,$ and $\tau, \ \lambda_1, \dots, \lambda_p \sim \mathcal{HC}(0,1)$, where $\mathcal{HC}$ denotes the half-Cauchy distribution. 
In this prior, $\tau$ corresponds to the global shrinkage and $\lambda_j$, to the local shrinkage. 
Then, inference is conducted through the posterior distributions, which are approximated through a Markov Chains Monte Carlo (MCMC) procedure. The estimator and its uncertainty (\ref{eq:estimator_general}) and (\ref{eq:uncertainty_general}) become
\begin{align*}
    &\widehat{\overline{Y}}_c = f_c \overline{y}_c^{(s)} + (1-f_c)\left(\frac{1}{L}\sum_{\ell=1}^L \widehat{\overline{Y}}_c^{(ns) (\ell)}\right),\\
    &\mathrm{PI}_{(1-\alpha)\%}\left[\widehat{\overline{Y}}_c\right] = f_c \overline{y}_c^{(s)} + (1-f_c)\left[\widehat{\overline{Y}}_{c, \mathrm{lower}_\alpha}^{(ns)}, \widehat{\overline{Y}}_{c, \mathrm{upper}_\alpha}^{(ns)}\right],
\end{align*}
where $\widehat{\overline{Y}}_c^{(ns) (\ell)} \sim \mathcal{N}\left(\overline{\bm{x}}_c^{(ns)^\top}\bm{\beta}^{(\ell)}, \sigma^{(\ell)^2}/(N_c-n_c)\right), \ \ell=1, \dots, L, \ c=1, \dots, M$, is the $\ell$th element of the MCMC posterior predictive sample, with $\bm{\beta}^{(\ell)}$ and $\sigma^{(\ell)}$ the $\ell$th elements in the MCMC samples. The $\alpha$-level empirical quantiles from the posterior predictive sample are denoted $\widehat{\overline{Y}}_{c, \mathrm{lower}_\alpha}^{(ns)}$ and $\widehat{\overline{Y}}_{c, \mathrm{upper}_\alpha}^{(ns)}$. \\

\section{Simulation study} 
\label{sec:Analyses}

%%%%%%%%%%%%%%%%%%%%%%%%%%%%%%%
%
% SIMULATIONS
%
%%%%%%%%%%%%%%%%%%%%%%%%%%%%%%%

This section presents two simulation studies to assess the performance of the proposed scaled SC procedure and to compare the four modelling methods. Section \ref{sec:Sim_scaledSC} focuses on the proposed scaled SC method that computes prediction intervals while relaxing the assumption of i.i.d. data points. In Section \ref{sec:Sim_comparison}, different generating models and sampling designs are studied to compare the four model selection methods described in Section \ref{sec:Model}.

Inference is performed in \texttt{R}. 
The random forests of $B=1000$ trees are trained using the \texttt{ranger} package \citep{wright2020ranger}. For each simulation scenario, the random forest hyperparameters are fixed after a cross-validation study of different values. The code to conduct the proposed scaled SC procedure for random forest estimates is available in appendix \ref{App:SC_Code}.
The LASSO method is conducted through the \texttt{glmnet} package, using the \texttt{cv.glmnet} function to define the optimal shrinkage penalty parameter.
The Bayesian inference is performed with the \texttt{NIMBLE} package \citep{nimble-article:2017}. Convergence of the MCMC chains is assessed through trace plots, effective sample sizes and the $\widehat{R}$ statistic \citep{gelman1992inference}. 

\subsection{Simulation study: scaled split conformal procedure}
\label{sec:Sim_scaledSC}

To assess the performance of the proposed scaled SC procedure, five simulation scenarios are considered. One finite population is created, and the different simulation scenarios correspond to the various sampling designs applied to that finite population. Let a finite population of $M=500$ areas of sizes $N_c, \ c=1, \dots, M,$ with $\min_c(N_c)=50$ and $\max_c(N_c)=500$. For $c=1, \dots, M,$ and $k=1, \dots, N_c$, the response variable has distribution $$y_{ck} \sim \mathcal{N}(9.5 + x_{1, ch} - x_{2, ch} + 2x_{3, ch} - x_{4, ch} + 2x_{5, ch} + x_{6, ch}, 1),$$ with 6 unit-level covariates, $x_1, \dots, x_6 \overset{i.i.d.}{\sim} \mathcal{N}(0,1)$. From the finite population, $R=500$ samples are drawn according to five sampling designs, which constitute the simulation scenarios:
\begin{enumerate}
    \item (Stratified) Select all $m=M=500$ areas and within each area, sample $n_c = 0.5N_c, \ c=1, \dots, m$ units;
    \item (Stratified) Select all $m=M=500$ areas and within each area, sample $n_c = 0.7N_c, \ c=1, \dots, m$ units;
    \item (One-stage) Sample $m = M/2$ areas and within each area, select all $n_c = N_c, \ c=1, \dots, m$ units;
    \item (Two-stage) Sample $m=M/2$ areas and within each area, sample $n_c = 0.5N_c, \ c=1, \dots, m$ units;
    \item (Two-stage) Sample $m=M/2$ areas and within each area, sample $n_c = 0.7N_c, \ c=1, \dots, m$ units.
\end{enumerate}
The proportion of sampled areas is higher in the stratified sampling designs, as all areas are selected. Hence, the areal-level inference is conducted on more data points in the first two scenarios than in the remaining three, and we expect any modelling method to perform better in these two scenarios. The one-stage and two-stage designs all yield $m=500$ areal-level responses. The difference between these last three scenarios is in the sampling fraction within areas. Out of the five simulation scenarios considered, the fourth one is the closest to the Ghanaian data analysed in Section \ref{sec:GAMA}.

For each simulation scenario and in each sample, the estimates described in equation (\ref{eq:estimator_general}) are computed using four methods: a linear model that includes the correct six covariates, a linear model that omits $x_4$, $x_5$ and $x_6$, a random forest method that considers all six covariates to grow the trees, and a linear LASSO model. The random forest hyperparameters are set after a cross-validation study as $mtry=2$ and $nodesize=5$. For each scenario, in each sample and for each modelling method, 50\%, 80\% and 95\% prediction intervals (\ref{eq:uncertainty_general}) are computed following the SC procedure and the proposed scaled SC procedure. These non-parametric methods may be applied to any modelling approach: a linear regression method as well as a machine learning method. The objective of this simulation study is to assess whether the proposed scaled SC method yields valid coverage rates of the prediction intervals.

\begin{figure}[!htb]
    \centering
    \includegraphics[width=\textwidth]{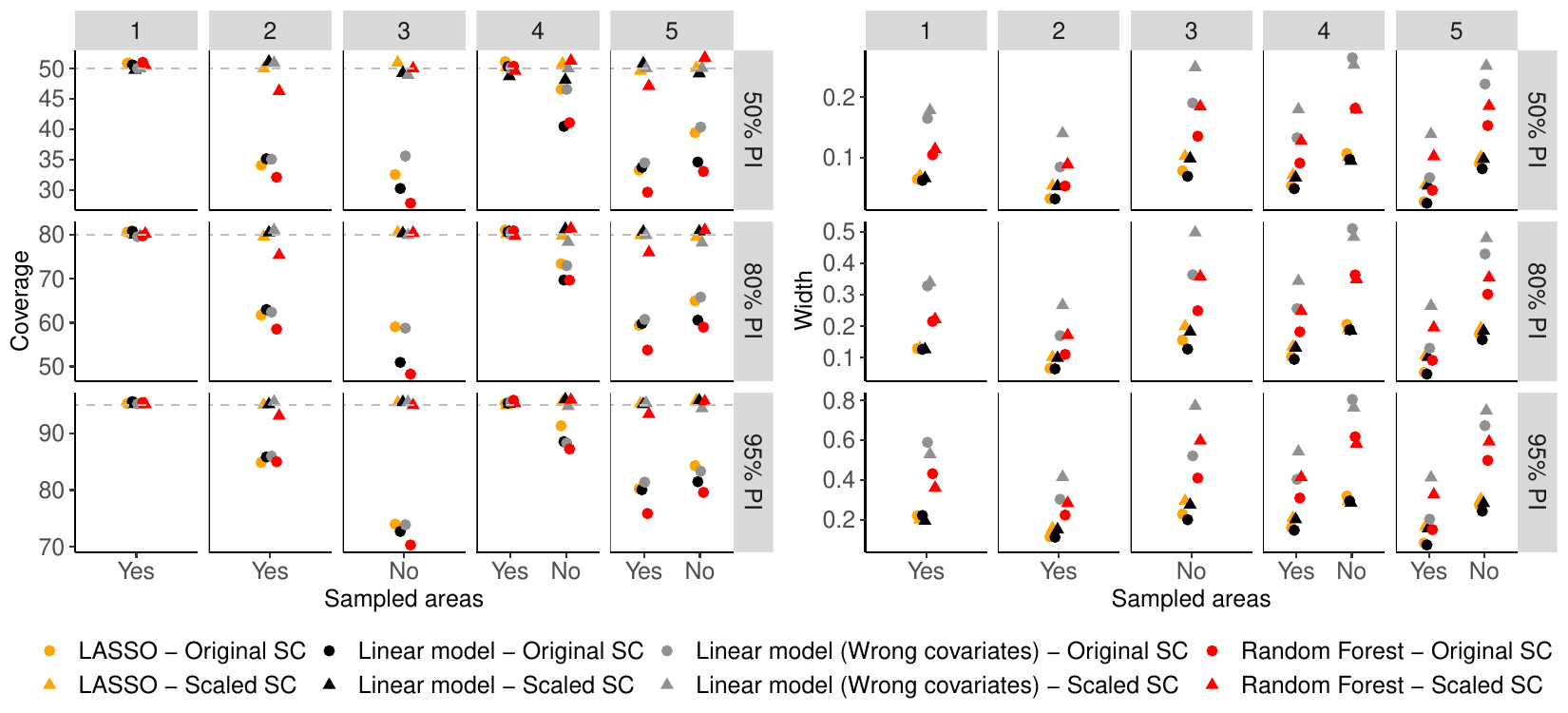}
    \caption{Coverages and widths of the prediction intervals (PI) obtained from the proposed scaled and original split conformal (SC) procedures for the four modelling methods and across the five simulation scenarios (1-5). Yes: coverages and widths across the sampled areas; No: coverages and widths across the non-sampled areas.}
    \label{fig:Sim_scaledSC_CovRange}
\end{figure}

All results are shown in Figure \ref{fig:Sim_scaledSC_CovRange}. The observed coverages for each scenario, method and interval level are shown in the left panel and the interval widths, in the right panel. The simulation scenarios are identified by their 1-5 number, as described above. Further, we differentiate the results for the sampled and non-sampled areas (Yes/No, respectively). Scenarios 1 and 2 only present results for sampled areas, as all areas are sampled in a stratified sample. Scenario 3 only shows the results for non-sampled areas because the target is exactly estimated in the sampled areas since all units are selected in a one-stage sample. Therefore, in the third scenario, we measure the predictions' uncertainty only in the non-sampled areas.

In the first scenario and for the sampled areas in the fourth scenario, $n_c$ and $N_c-n_c$ are equal. Therefore, the data points are i.i.d. and the SC and proposed scaled SC approaches are the same. In these scenarios, both methods yield the right coverages of the prediction intervals, regardless of the modelling method. In terms of interval widths, the linear model with incorrect set of covariates leads to the widest intervals under both SC procedures. The linear model with correct covariates and the LASSO method yields the narrowest intervals regardless of the SC method. 

For all other sampling schemes, $n_c \neq N_c - n_c$. Therefore, the sampled and non-sampled means, $\overline{Y}_c^{(s)}$ and $\overline{Y}_c^{(sn)}$, are not identically distributed, with differently scaled variances, $\sigma^2/n_c$ and $\sigma^2/(N_c-n_c)$, respectively. In these cases, the original SC intervals obtained for all four modelling methods do not attain the right coverages. The original SC leads to under-coverage of the prediction intervals. On the other hand, the proposed scaled SC procedure produces prediction intervals with the right coverages, regardless of the interval level and modelling method. In particular, when fitting a linear model, with the LASSO constraint or with and without the right mean structure, the scaled SC intervals have exactly the right coverages. When modelling with a non-parametric random forest approach, the scaled SC prediction intervals show a slight error in the coverage rate. Specifically, in scenarios 2 and 5, the random forest shows slight under-coverage. 

In terms of interval width, the proposed scaled SC intervals tend to be a little wider than the original SC ones, for all interval levels. Regardless of the simulation scenario, the SC intervals and proposed scaled SC intervals obtained for the random forest estimates tend to be narrower than the ones obtained for the linear estimates using the incorrect set of covariates.

\subsection{Simulation study: prediction methods comparison}
\label{sec:Sim_comparison}

To compare the performance of the random forest and the LASSO methods to the frequentist forward variable selection and the Bayesian shrinkage approaches, as described in Section \ref{sec:Model}, a simulation study is conducted considering three generating models for the outcome and five sampling designs. Three finite populations of $M=1000$ areas of sizes $N_c$ with $\min_c(N_c)=50$ and $\max_c(N_c)=500$ are created as follows, for $c=1, \dots, M$ and $k=1, \dots, N_c$:

\begin{enumerate}
    \item[A.] $y_{ck} \sim \mathcal{N}\left(20 + \bm{x}_{ck}^\top\bm{\beta}, 0.5^2\right),$ where the covariates are such that $\bm{x}_{ck} \sim \mathcal{N}_{100}(\bm{0},\bm{I})$ and  with coefficients $\bm{\beta}^\top=(1, -1, 2, -1, 2, 1, 2, 1, -1, 1, 0, \dots, 0)$;
    \item[B.] $y_{ck} \sim \mathcal{N}\left(20 + \bm{x}_{ck}^\top\bm{\beta}, 0.5^2\right),$ where the covariates are such that $\bm{x}_{ck} \sim \mathcal{N}_{100}(\bm{0},\Sigma_x),$ with $\Sigma_x = \begin{bmatrix}1 & 0.5 & \dots & 0.5 \\
    0.5 & 1 & \dots & 0.5 \\
    \vdots & \vdots & \ddots & \vdots\\
    0.5 & 0.5 & \dots & 1\end{bmatrix},$ and $\bm{\beta}^\top=(1, -1, 2, -1, 2, 1, 2, 1, -1, 1, 0, \dots, 0)/10$;
    \item[C.] $y_{ck} \sim \mathcal{N}\left(x_{1,ch}^2 + \exp\left(x_{2,ch}^2\right), 0.3\right),$ with covariates $x_{j,c} \sim \mathcal{U}(-1,1),$ $j=1, \dots, 100$.
\end{enumerate}

Populations A and B assume a linear relationship between the outcome and the first 10 covariates. In scenario B, however, the strength of the association is weak and the covariates are correlated. Population C is inspired by \cite{scornet2017tuning} and assumes a non-linear relationship between the outcome and covariates. Throughout this simulation study, areas are indiscriminately termed ``areas'' or ``EAs''. From each finite population, $R=100$ samples are drawn following the two sampling schemes:
\begin{enumerate}
    \item (Stratified) Select all $m=M=500$ areas and within each area, sample $n_c = 15, \ c=1, \dots, m$ units;
    \item (Two-stage) Sample $m=M/2$ areas and within each area, sample $n_c = 15, \ c=1, \dots, m$ units.
\end{enumerate}

These simulation scenarios were motivated by the Ghanaian data analysed in Section \ref{sec:GAMA}, where only 15 households were sampled within the selected areas. Additional sampling designs are considered in Appendix \ref{App:Sim}, with a higher number of sampled units within the selected areas. 
For each scenario, the estimates and their prediction intervals are computed as described in Section \ref{sec:Model}. Further, for each scenario, the estimates and their uncertainty are also computed assuming anonymised EAs. In this context, the modelling methods are trained on the sample and predictions are obtained ignoring which areas have been sampled, that is, assuming $f_c=0, \ c=1, \dots, M,$ at the prediction stage. Once again, this study with anonymised EAs is run because the available Ghanaian data that is analysed in Section \ref{sec:GAMA} does not identify the sampled EAs. The random forest hyperparameters are set following a cross-validation study conducted for each simulation scenario. A random forest with hyperparameters set to $(mtry, nodesize)=(10,5)$ is found to perform the best for both sampling schemes in populations A and B. In population C, we set $(mtry, nodesize)=(70,200)$ and $(mtry, nodesize)=(70,9)$, for the stratified samples and the two-stage samples, respectively. In all scenarios, the Bayesian approach runs through a MCMC procedure with two chains of 5,000 iterations, which include a burn-in period of 2,500 iterations. We find out that with these values, practical convergence was attained, as assessed by the trace plots, effective sample sizes and $\widehat{R}$ statistics \citep{gelman1992inference}. For a particular finite population and sampling scheme, running the random forest over 100 replicates takes 55 minutes, while the LASSO takes 4 minutes, the forward method takes 7 minutes and the Bayesian approach, 2.5 hours.

The methods' performances are compared through the mean absolute bias, mean squared error (MSE), coverages of 50\%, 80\% and 95\% prediction intervals and their proper interval score \citep{gneiting2007strictly}. Let $\left[L,U\right]$ be the $(1-\alpha)\%$ prediction interval for a quantity $\theta$ to estimate. The proper score is defined as $(U-L) + 2/\alpha\left[(L-\theta)\mathds{1}_{L>\theta} + (\theta-U)\mathds{1}_{U<\theta}\right]$. Smaller values of the interval proper scores are preferred, indicating narrow intervals and average close to the nominal.
Additionally, we extract which covariates have been selected from each method. Note that in the Bayesian framework, a covariate is said to be selected when its coefficient's posterior 95\% credible interval does not include 0. For the random forest approach, when the $p$-value related to a variable's importance \citep{altmann2010permutation} is smaller than $0.05$, said variable is deemed selected. The variable importance is computed based on results from random forests fitted with permutations of the set of covariates.

\begin{figure}[!htb]
    \centering
    \includegraphics[width=\textwidth]{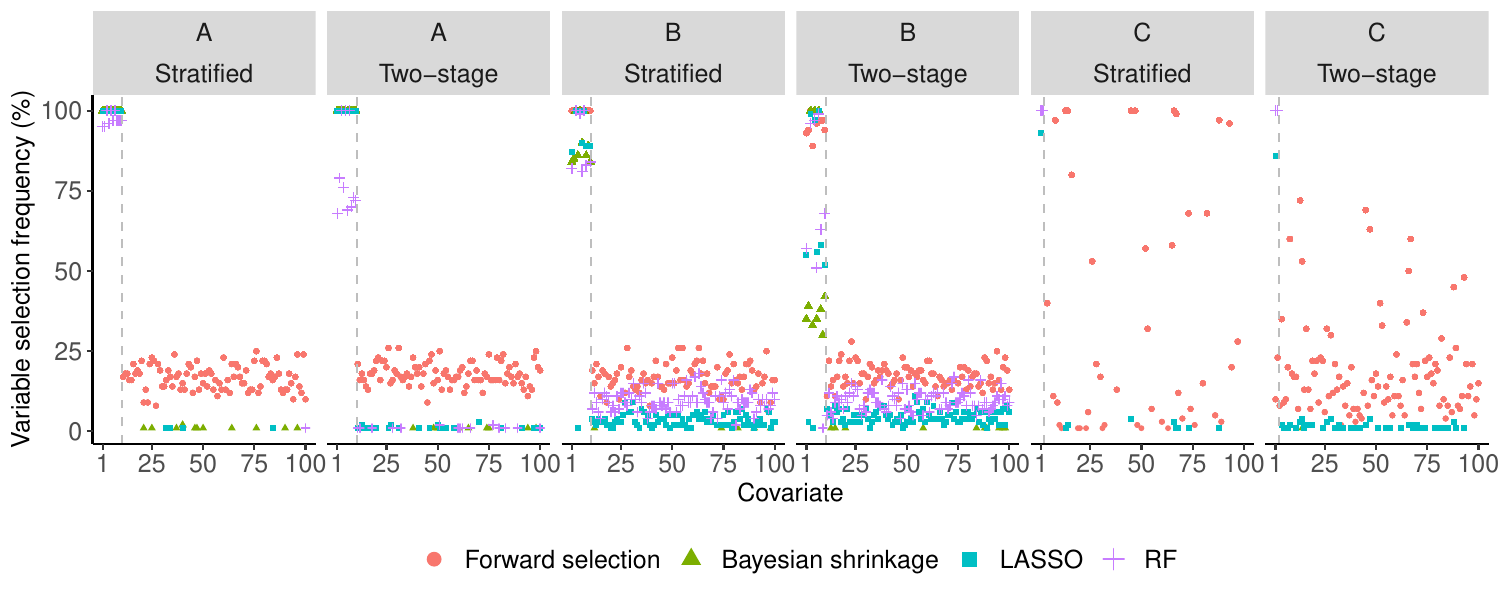}
    \caption{Covariate selection frequency for each method across the 6 simulation scenarios. Left of the vertical dashed line: true covariates used in the generating models.}
    \label{fig:Sim_comparison_Covariates}
\end{figure}

Figure \ref{fig:Sim_comparison_Covariates} shows the selected covariates by each method for each finite population and sampling design. When the association is linear between the covariates and the outcome (A and B), regardless of the sampling design, the forward approach tends to adequately select the true auxiliary information. However, it also tends to select irrelevant variables. Each unimportant covariate is selected about 25\% of the time by the forward method. In population A, the LASSO and Bayesian approaches also select the right covariates 100\% of the time, while almost never including redundant covariates. When the association is weak, the LASSO and Bayesian methods tend to miss the right covariates 20\%--50\% of the time, depending on the sampling design. In scenarios A and B, the random forest method misses the right covariates 10\%--50\% of the time, while it always captures the correct set when the association is non-linear. The LASSO selects 1 out of 2 correct covariates about 80\% of the time in this third population. Both the forward and Bayesian approaches miss the correct set of covariates in scenario C almost 100\% of the time and include irrelevant variables. 

\begin{figure}[!htb]
    \centering
    \includegraphics[width=\textwidth]{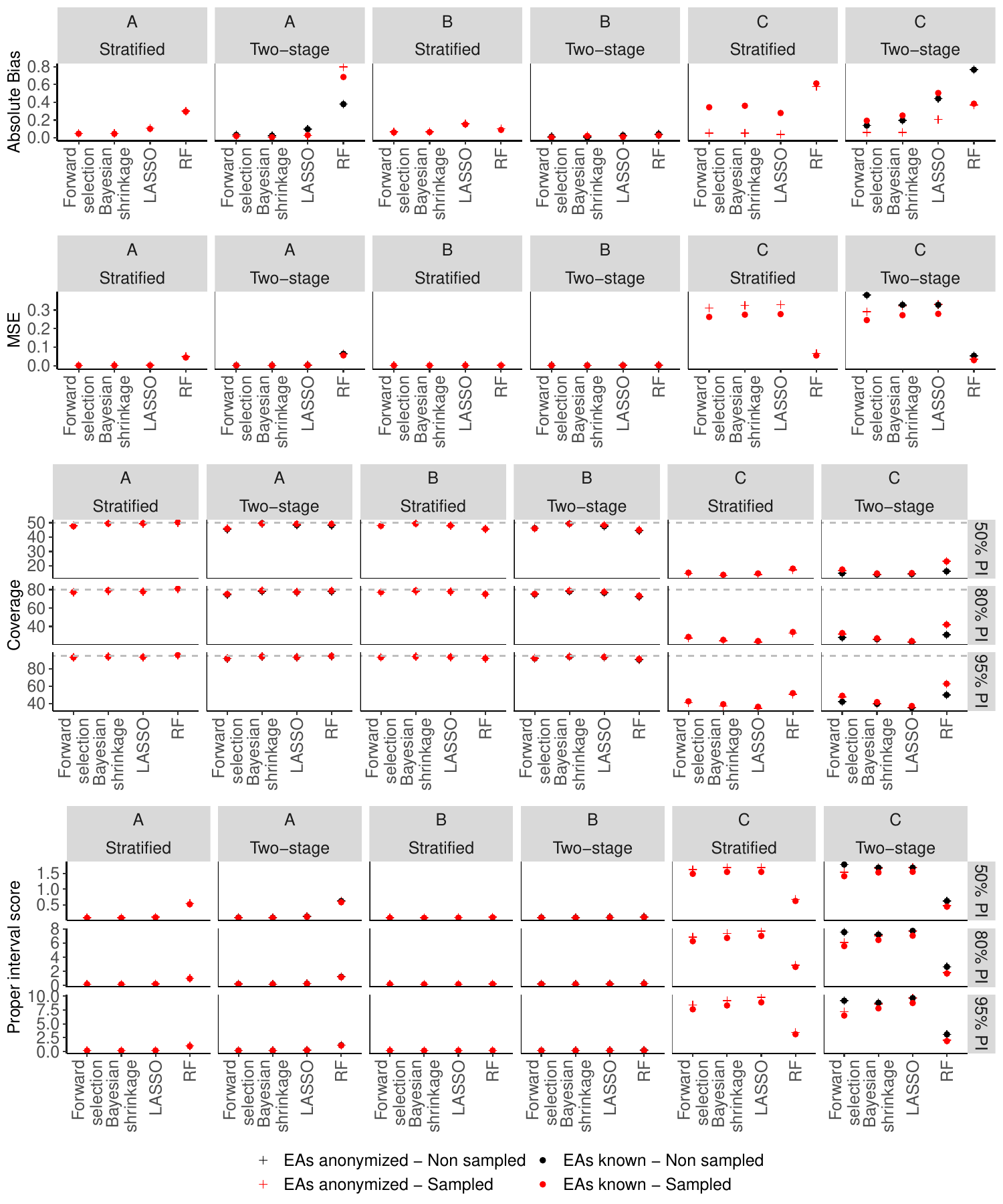}
    \caption{Mean absolute bias, MSE, coverages and proper scores of the prediction intervals, obtained for each method across the 6 simulation scenarios. RF: Random forest approach.}
    \label{fig:Sim_comparison_Results}
\end{figure}

Figure \ref{fig:Sim_comparison_Results} shows the absolute biases multiplied by 100, MSEs, prediction intervals' coverages and proper scores for all methods, generating models (A-C) and sampling schemes. The results for the sampled and non-sampled areas are differentiated through the red and black symbols, respectively. Only results for the sampled EAs are produced for the stratified sampling design, as all areas are sampled. The results assuming the anonymised EAs are distinguished from the ones in which we know which areas have been sampled by the circle and cross symbols, respectively.

For all performance measures, the four modelling approaches yield similar results when it is known and unknown which areas have been sampled. For example, in population C with a two-stage sampling design and regardless of the modelling method, the MSE results over the anonymised sampled EAs are not worse than the results over the non-sampled EAs. This result is reassuring as for analysing the Ghanian data, where the sampled EAs are anonymised. 

In terms of bias, all methods are virtually unbiased with mean absolute biases between 0 and 0.8, regardless of the population and sampling design. The random forest tends to yield slightly higher mean absolute biases, compared to the forward, LASSO and Bayesian methods. As expected, the mean absolute biases are higher for all modelling methods in the non-linear scenario (C). Interestingly, the random forest, which always selects the right covariates in scenario C (see Figure \ref{fig:Sim_comparison_Covariates}), yields larger biases than the other three methods that miss the right set of covariates almost 100\% of the time. 

In terms of MSE, there does not seem to be a difference between the LASSO, forward and Bayesian methods, for all populations and sampling schemes. These three methods, which fit linear models, yield slightly smaller MSEs than the random forest approach when the association between the covariates and the outcome is strongly linear (A). For scenario C, however, the random forest produces smaller MSEs than the three linear modelling approaches. The random forest method divides the other three modelling methods' MSEs by a factor of 3 in population C, regardless of the sampling scheme.

The prediction intervals computed for all four methods in each sampling scheme for populations A and B yield the right coverages. These intervals are wider for the random forest method in population A, for both sampling designs, as deduced from the proper interval scores. When the relationship between the outcome and the covariates is non-linear (C), we observe that all four modelling methods yield under-coverage in both sampling designs. The random forest method, which accommodates a non-linear relationship, leads to prediction intervals with slightly higher coverage rates than the other three methods, but still misses the right rates by about 30\%. Note, however, that the random forest approach produces prediction intervals with smaller proper scores than the other three modelling methods.

\section{Areal log consumption prediction in the Greater Accra Metropolitan Area}
\label{sec:GAMA}

%%%%%%%%%%%%%%%%%%%%%%%%%%%%%%%
%
% DATA APPLICATION 
%
%%%%%%%%%%%%%%%%%%%%%%%%%%%%%%%

In this section, the four modelling methods described in Section \ref{sec:Model} are applied to the data for the Greater Accra Metropolitan Area (GAMA) in Ghana. Using the sixth GLSS and the 2010 Ghanaian census, a complete map of the mean equivalised consumption (in the log scale) is produced across the $M=5019$ enumeration areas (EAs), for each method. Note that in the household survey, only $m=136$ EAs have been sampled. To provide estimates in the missing areas, the response values are modelled using the $p=174$ auxiliary variables which are measured in both available datasets. 

For the random forest approach, a cross-validation study on the survey data was run to set the hyperparameters to $B=1000$ trees grown with $mtry = 25$ and $nodesize=3$. 
The Bayesian approach required two MCMC chains of 100,000 iterations, including a burn-in period of 50,000, and a thinning factor of 15. Convergence was attained as assessed by the trace plots, effective sample sizes and $\widehat{R}$ statistics. 

\begin{figure}[!htb]
    \centering
    \includegraphics[width=\textwidth]{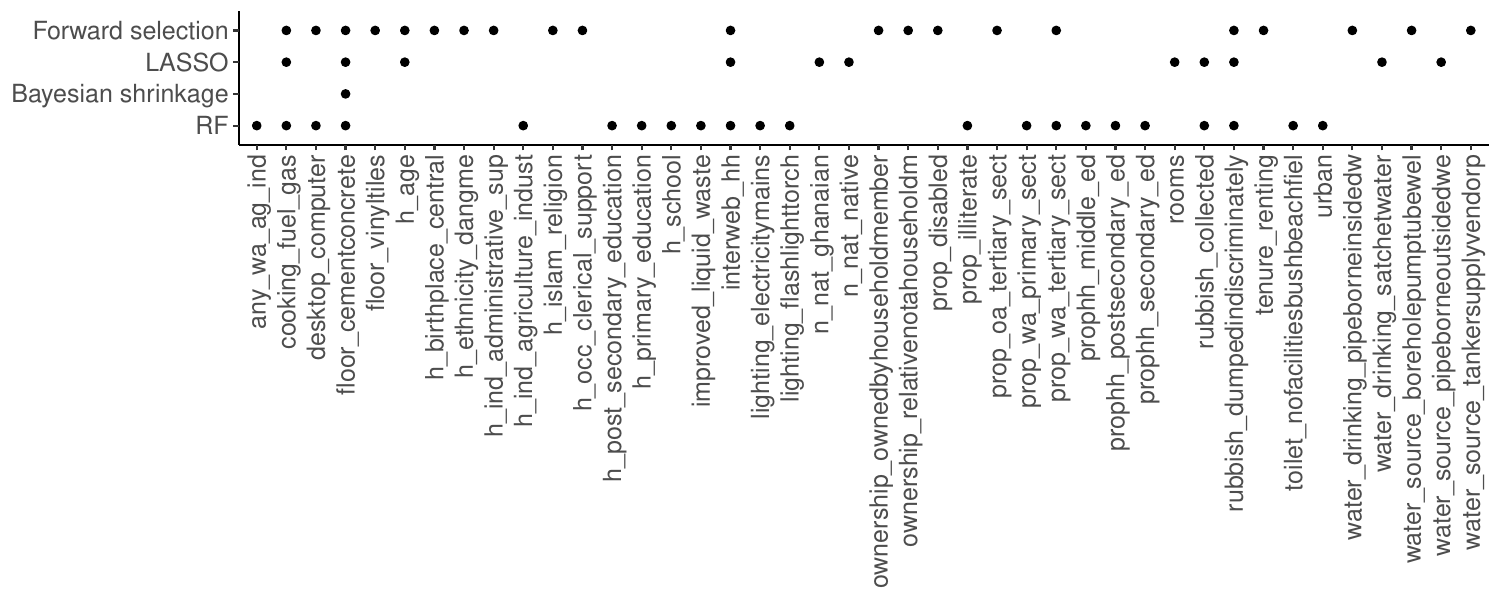}
    \caption{Selected covariates for each method when modelling the log equivalised consumption in GAMA.}
    \label{fig:GAMA_Covariates}
\end{figure}

\cite{wakefield2020small} point out the importance of including the design variables in model-based small area estimation methods. To that end, the urban indicator, which corresponds to the sample strata, is added to all four modelling methods. In the forward selection approach, this inclusion means that the urban indicator is added to the vector of selected covariates, even if it was not selected in the first step. In the Bayesian shrinkage and LASSO methods, it means there is no shrinkage applied to the regression coefficient that corresponds to the urban indicator. Finally, in the random forest approach, it means that the urban indicator is part of the variables considered for each covariate split. Figure \ref{fig:GAMA_Covariates} presents the covariates that were selected by each method. Despite all methods including the urban indicator, only the random forest finds it relevant with a $p$-value for its variable importance smaller than 0.05. Additionally, Figure \ref{fig:GAMA_Covariates} shows that the horseshoe prior leads to only one variable whose coefficient's posterior 95\% credible interval does not include 0. The LASSO approach selects about 6\% of the available covariates (11 variables), while the forward method and random forest methods select more than 12\% of the variables (21 and 22, respectively). The variable indicating whether a household's floor is made of cement or concrete is selected by all four methods. 

\begin{figure}[!htb]
    \centering
    \includegraphics[width=\textwidth]{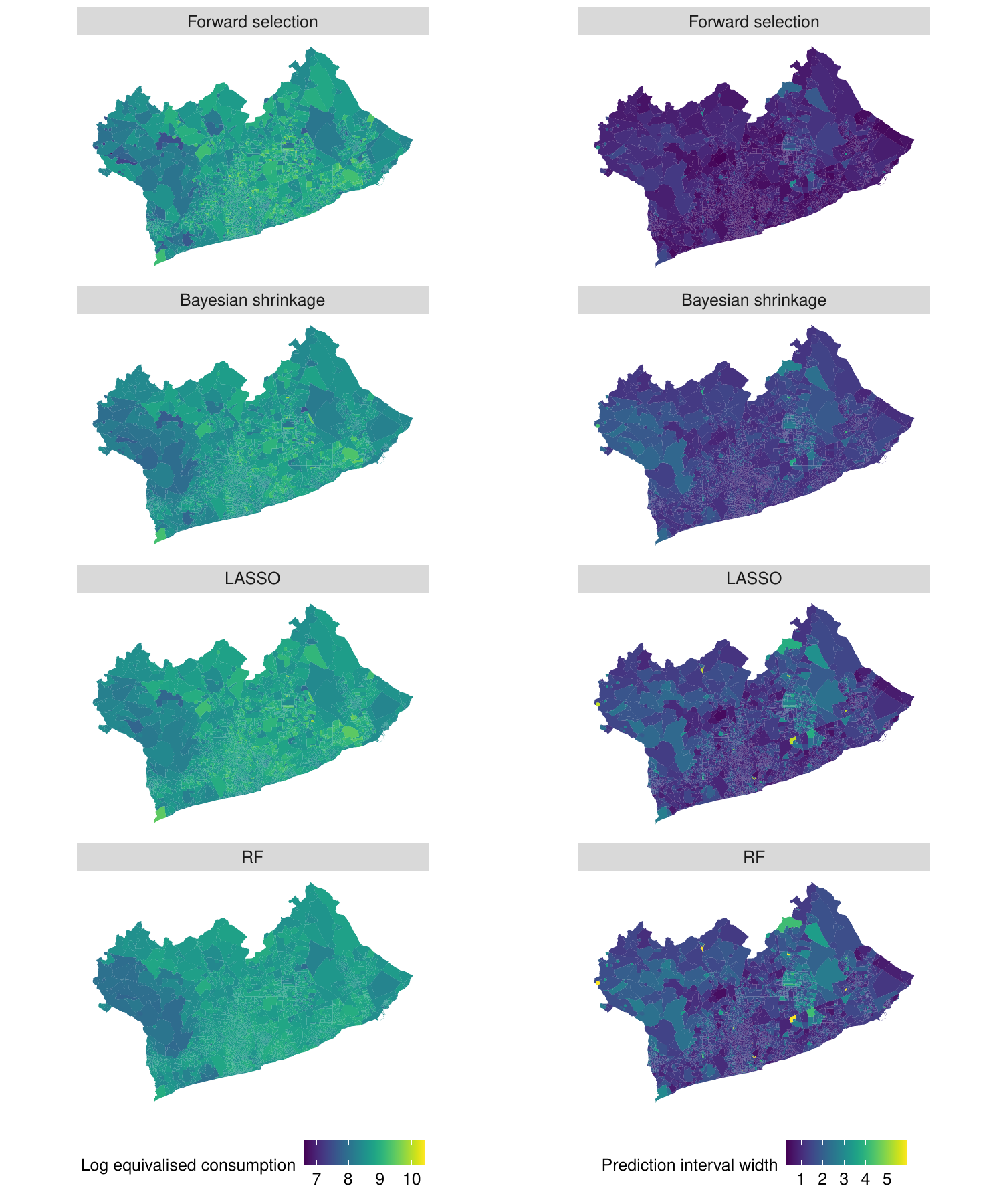}
    \caption{Estimated mean log equivalised consumption in the GAMA EAs (Left) and widths of the corresponding 95\% prediction intervals (Right) obtained from each modelling method. RF: Random forest.}
    \label{fig:GAMA_pred}
\end{figure}

\begin{figure}[!htb]
    \centering
    \includegraphics[width=\textwidth]{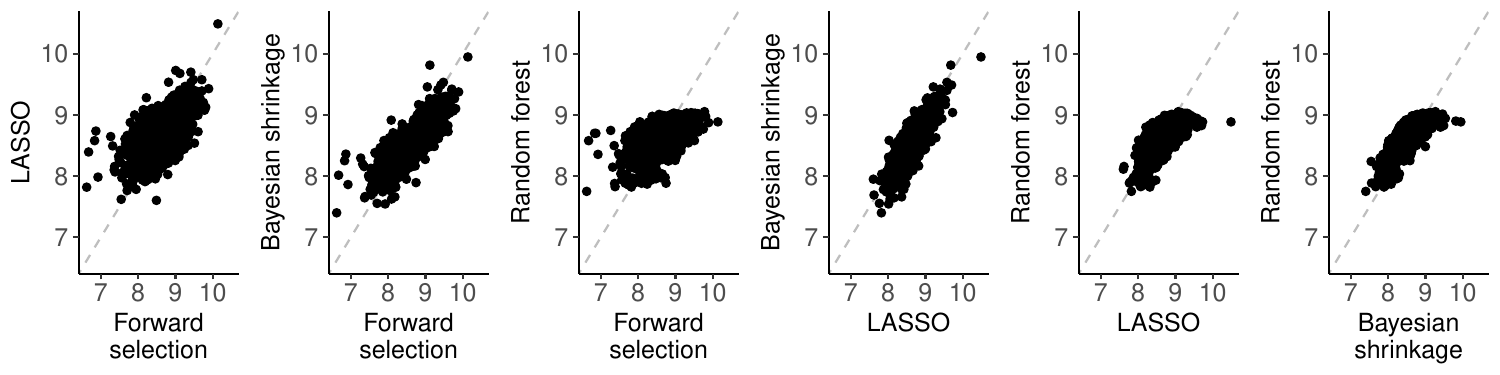}
    \caption{Pairwise comparison of the areal estimates obtained from each of the four methods: forward selection, LASSO, Bayesian shrinkage and random forest.}
    \label{fig:GAMA_ComparePred}
\end{figure}

\begin{figure}[!htb]
    \centering
    \includegraphics[width=\textwidth]{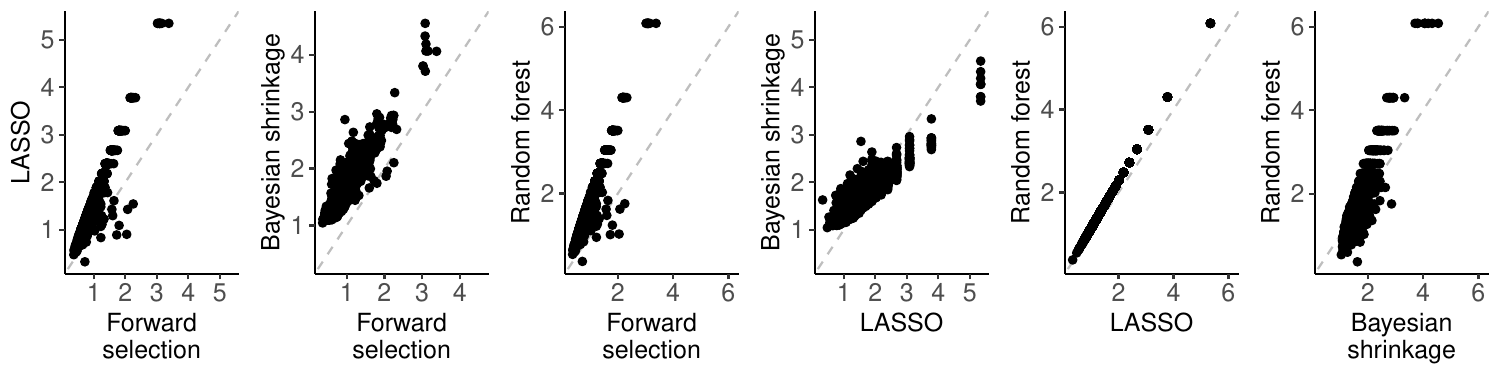}
    \caption{Pairwise comparison of the areal prediction interval widths obtained from each of the four methods: forward selection, LASSO, Bayesian shrinkage and random forest.}
    \label{fig:GAMA_CompareWidths}
\end{figure}

Figure \ref{fig:GAMA_pred} shows the mean log consumption areal estimates and their 95\% prediction intervals' widths for each of the four methods. Among the four methods, The random forest approach yields the most homogeneous point estimates across the EAs. This can further be seen in Figure \ref{fig:GAMA_ComparePred} which compares the predictions obtained using each method for each EA. The prediction interval widths are shown across the EAs in Figure \ref{fig:GAMA_pred} and compared between the modelling methods in pairwise scatter plots in Figure \ref{fig:GAMA_CompareWidths}. The prediction intervals computed for the linear approach with forward variable selection are the narrowest. As expected, the widths of the intervals obtained through the proposed scaled SC approach for the random forest and LASSO predictions behave similarly. The widths are of the form $(1-f_c)\times 2 \times d_\alpha/\sqrt{N_c-n_c}$, where $d_\alpha$ is the only quantity that differs between the LASSO and random forest approaches. Note that in this analysis, the scaled SC procedure divides the dataset into two halves, consequently computing the necessary residuals and quantile based on only $m/2=68$ data points. 

\begin{figure}[!htb]
    \centering
    \includegraphics[width=0.7\textwidth]{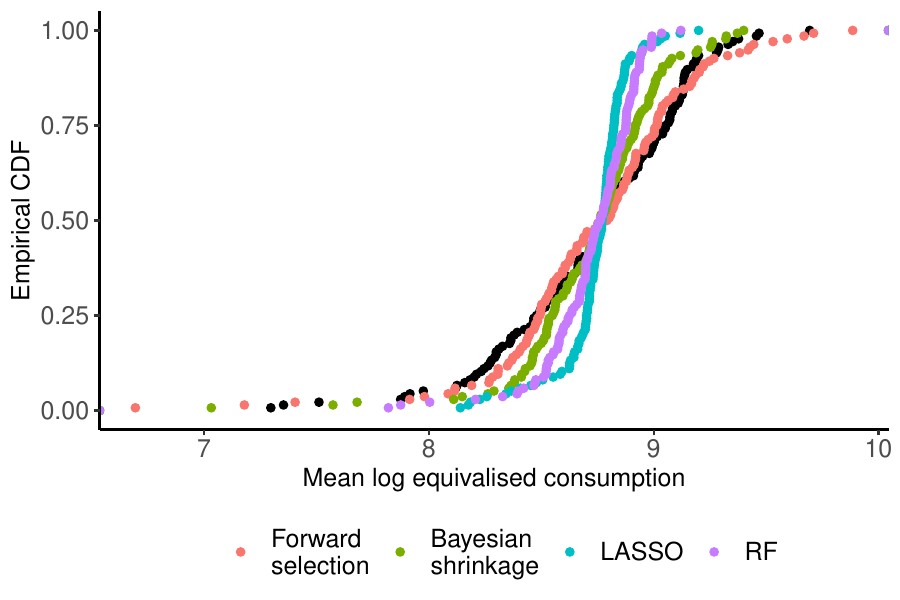}
    \caption{Empirical CDFs obtained for each method in the 8-fold cross-validation study on the GAMA sample. Black: empirical CDF for the observed sampled means.}
    \label{fig:CV_GAMA_CDF}
\end{figure}

Finally, to determine which method performs the best in this particular data application, an 8-fold cross-validation study is conducted. The 136 sampled EAs are divided into 8 rural EAs and 128 EAs. Hence, in this 8-fold cross-validation study, 17 EAs are removed from the sample at a time (1 rural and 16 urban EAs), the four methods are fitted on the remaining 119 EAs and predictions are obtained for the 17 removed ones. The four methods are compared in terms of mean absolute bias, MSE, coverages and proper scores of the 50\%, 80\% and 95\% prediction intervals in Table \ref{tab:CV_GAMA}. The Bayesian shrinkage approach performs the best among the four methods we consider, yielding the smallest bias, MSE and interval scores and reaching the right coverage rates of the prediction intervals. On the other hand, the prediction intervals obtained for the forward selection approach lead to significant undercoverage. 
To further compare the performance of the four modelling approaches, the empirical cumulative distribution functions (CDFs) of the point estimates obtained by each method are shown in Figure \ref{fig:CV_GAMA_CDF}, alongside the empirical CDF of the observed Ghanaian sampled means. Figure \ref{fig:CV_GAMA_CDF} shows that the forward selection method performs the best when estimating the empirical distribution of the mean log equivalised consumption. The Bayesian shrinkage method is the second best method to estimate the empirical distribution of the mean log equivalised consumption. 
A cross-validation study was also conducted where the four methods were fitted without forcing the inclusion of the urban indicator. The results are not shown in this paper as the performance of the four methods in terms of bias, MSE, coverage and proper score of the prediction intervals were similar to the ones shown in Tables \ref{tab:CV_GAMA} and \ref{fig:CV_GAMA_CDF}, obtained including the urban indicator, for each modelling method.
The Bayesian shrinkage method considered in this paper consists in applying the horseshoe prior to the regression coefficients. Other priors could have been considered, such as a Bayesian ridge prior. A cross-validation study was conducted with the Bayesian ridge approach for the GAMA sample. Because the results were similar to the ones shown in Tables \ref{tab:CV_GAMA} and \ref{fig:CV_GAMA_CDF}, obtained with the horseshoe prior, in terms of bias, MSE, coverage and proper score of the prediction intervals, they are not presented in this paper.

\begin{table}[!htb]
	\centering
	\begin{tabular}{l cccccccc}
		\toprule
		& Absolute & \multirow{2}{*}{MSE} & \multicolumn{3}{c}{PI Coverage} & \multicolumn{3}{c}{Proper interval score}   \\
		& Bias & & 95\% & 80\% & 50\% & 95\% & 80\% & 50\% \\
		\midrule
		Bayesian shrinkage & 0.244 & 0.086     &   94.1    &    80.9     &   48.5   &  1.33   &  1.91    & 3.86 \\ 
		Forward selection & 0.975 & 0.168   &     72.1     &   52.2     &   28.7   &  3.55  &   5.35  &   8.03 \\
		LASSO & 0.965 & 0.133   &     91.9     &   76.5      &  49.3   &  1.75  &   2.48  &   4.65 \\
		Random forest & 0.516 & 0.097      &  91.9     &   79.4  &     50.0  &     1.61  &   2.08  &   4.16 \\
		\bottomrule
	\end{tabular}
 \caption{Mean absolute bias, MSE, coverages and proper scores of the 50\%, 80\% and 95\% prediction intervals, obtained for each method in the 8-fold cross-validation study on the GAMA sample.}
 \label{tab:CV_GAMA}
\end{table}

\begin{figure}[!htb]
    \centering
    \includegraphics[width=\textwidth]{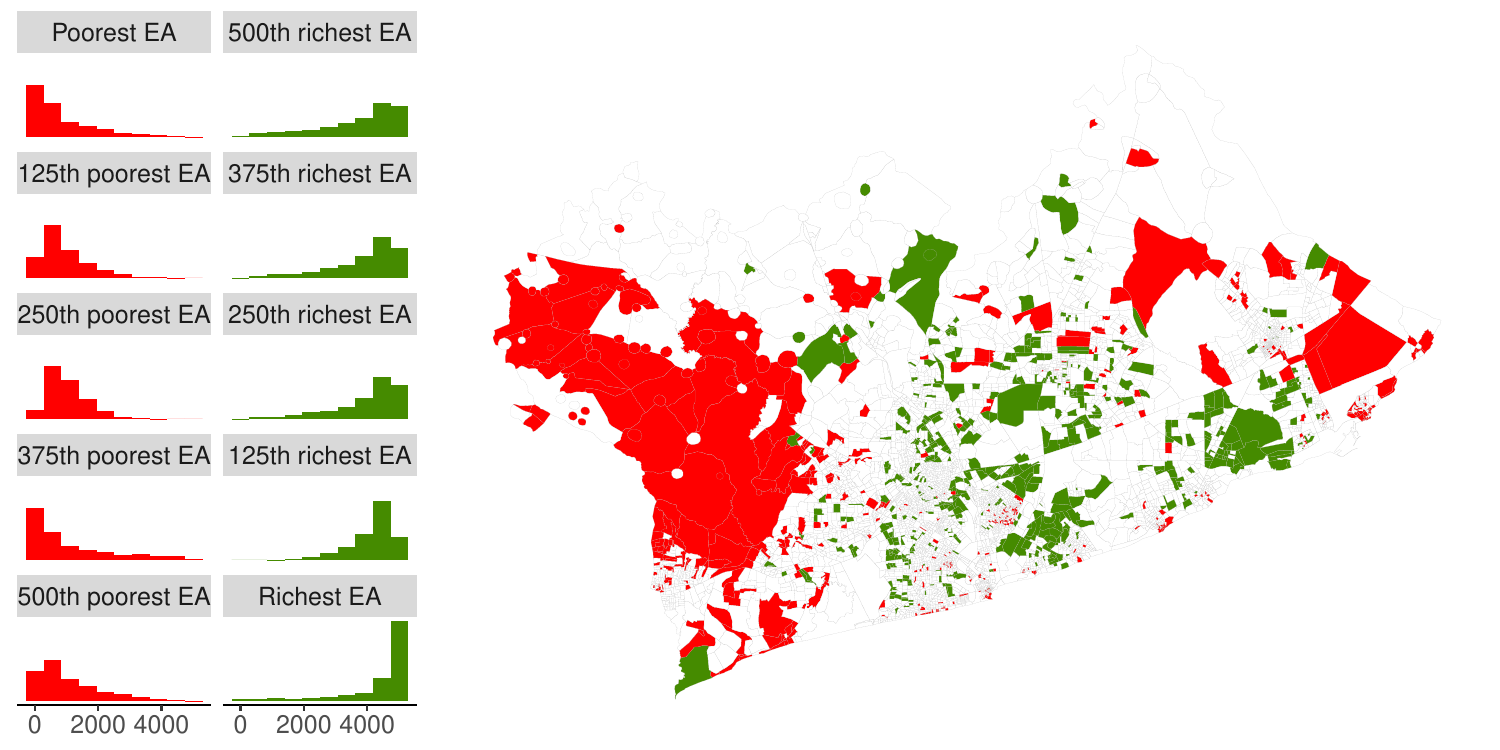}
    \caption{Left: Histograms of the posterior ranking distributions of 5 of the 10\% poorest EA's (left column, red) and 5 of the 10\% richest EA's (right column, green), as estimated from the MCMC samples obtained for the Bayesian shrinkage approach. Right: Map highlighting the 500 poorest EA's (red) and the 500 richest EA's (green). There are a total of 5019 EAs in the study region.}
    \label{fig:GAMA_ranked}
\end{figure}

In the original scale, on average, we find that the Bayesian shrinkage consumption estimates among the richest 10\% are 2.3 times bigger than the ones among the poorest 10\%. We also find that the 92\% urban EAs are not uniformly distributed across the estimated consumption deciles: there are only 79\% urban EAs among the poorest 10\%, versus 91\% among the richest 10\%. Following \cite{dong2021modeling}, to identify the EAs where interventions should be prioritized, we rank the EAs from poorest to richest, based on the Bayesian shrinkage point estimates. In particular, in this Bayesian framework, we obtain each EAs posterior ranking distribution, by ranking the point estimates at each MCMC iteration. Figure \ref{fig:GAMA_ranked} shows the posterior ranking distributions for 5 of the 10\% poorest EAs and 5 of the 10\% richest EAs. Additionally, the right-hand side of Figure \ref{fig:GAMA_ranked} maps the 10\% poorest and richest EAs. We find that the Greater Accra South district, which corresponds to the western EAs in Figure \ref{fig:GAMA_ranked}, gathers most of the poorest EAs, while the Accra Metropolitan Area district, which corresponds to the southern EAs in Figure \ref{fig:GAMA_ranked}, is the richest. Figure \ref{fig:GAMA_ranked} further shows that the 500 poorest EAs' ranking distributions overlap, which seems to indicate that there is a need to intervene in the poorest 500 EAs.

\section{Discussion}
\label{sec:Discussion}

%%%%%%%%%%%%%%%%%%%%%%%%%%%%%%%
%
% CONCLUSION
%
%%%%%%%%%%%%%%%%%%%%%%%%%%%%%%%

In this paper, we compare four methods that perform variable modelling to estimate area-level means of a variable of interest. Throughout, the areas correspond to the sampling clusters. The methods are area-level model-based small area prediction procedures used to obtain areal estimates and their uncertainties. First, a random forest approach models the outcome values. By construction, auxiliary variables are selected when partitioning the response values through covariate splits. Second, in the frequentist framework, a LASSO method selects covariates by shrinking irrelevant regression coefficients towards 0. Then, in the frequentist framework, a forward variable selection approach with the AIC as the optimisation criterion is considered. Finally, in the Bayesian framework, the horseshoe prior, which shrinks the coefficients towards zero, \textit{a priori}, is assumed for the regression coefficients.

Further, a modification of the split conformal (SC) procedure to compute prediction intervals is proposed. The SC algorithm \citep{lei2018distribution} estimates prediction intervals with no specific distribution assumption for the data. However, the data are assumed to be identically and independently distributed i.i.d.. The proposed scaled SC procedure relaxes the assumption that the data are identically distributed. Specifically, the proposed algorithm allows the data points to have variances of different scales. This proposed scaled SC procedure allows inference to be conducted for the random forest and the LASSO estimates.

A first simulation study assesses the performance of the proposed scaled SC method compared to the original SC procedure. It is found that when the data points are i.i.d., both procedures perform similarly, regardless of the modelling method. In the simulation scenarios where the number of sampled units is not equal to the number of non-sampled units ($n_c \neq N_c-n_c$), the variances are scaled differently, $\sigma^2/n_c \neq \sigma^2/(N_c-n_c)$. Hence, the SC procedure does not yield the appropriate coverage rates for the prediction intervals in these scenarios. The proposed scaled SC method corrects the under-coverage in all the simulation scenarios that were considered. 

The four variable selection methods under study are compared in an additional simulation study. When data are generated from a linear model, the methods that assume normality yield smaller biases and MSEs than the random forest approach. All modelling methods, however, lead to adequate prediction interval coverages. The random forest method performs better in terms of MSE when the data are generated from a non-linear model. All methods yield under-coverage when few units are selected within the sampled areas in this complex population.

In the sixth Ghana Living Standards Survey, from 2012--2013, the log equivalised consumption is measured at the household level in a small fraction of the areas (EAs) within the Greater Accra Metropolitan Area (GAMA), alongside 174 auxiliary variables. The same auxiliary information is recorded for all the GAMA EAs in the 2010 Ghanaian census. Using both datasets and the four EA-level method-based approaches, areal estimates of the mean log equivalised consumption are computed for all EAs in the GAMA. Additionally, prediction intervals are computed for all EA estimates to measure their uncertainties. The LASSO and forward variable selection methods select more than 10\% of the auxiliary variables, while the Bayesian horseshoe model yields posterior credible intervals that do not include 0 for only one coefficient. The random forest procedure estimates a smoother map of the mean log consumption than the other three approaches. A cross-validation study conducted on the sample data shows that the Bayesian shrinkage method performs the best, among the four methods considered, on this particular dataset.

Finally, in this paper, before fitting random forests to the different datasets, cross-validation studies were run to help set the hyperparameters. These hyperparameters are the number of regression trees included in the forest, the number of variables considered at each step when growing the trees, and the final node sizes. This step remains to be improved: as other hyperparameters could have led to better performing random forests. For further discussion on the selection of random forest hyperparameters; see e.g., \cite{mcconville2019automated, dagdoug2021model}. On the other hand, the proposed scaled SC procedure used to compute prediction intervals for the random forest and LASSO estimates relies on an equal split of the data points to grow a forest and compute prediction errors. In the data application of this paper, this implies that the prediction interval limits are based on 68 data points. This partition, suggested by \cite{lei2018distribution} for the original SC algorithm, could be revisited to attempt to narrow down the resulting intervals.

\bibliographystyle{apalike} % Style BST file (imsart-number.bst or imsart-nameyear.bst)
\bibliography{biblio}       % Bibliography file (usually '*.bib')

%% or include bibliography directly:
% \begin{thebibliography}{}
% \bibitem{b1}
% \end{thebibliography}

%%%%%%%%%%%%%%%%%%%%%%%%%%%%%%%%%%%%%%%%%%%%%%
%% Single Appendix:                         %%
%%%%%%%%%%%%%%%%%%%%%%%%%%%%%%%%%%%%%%%%%%%%%%
%\begin{appendix}
%\section*{???}%% if no title is needed, leave empty \section*{}.
%\end{appendix}
%%%%%%%%%%%%%%%%%%%%%%%%%%%%%%%%%%%%%%%%%%%%%%
%% Multiple Appendixes:                     %%
%%%%%%%%%%%%%%%%%%%%%%%%%%%%%%%%%%%%%%%%%%%%%%
\begin{appendix}

\section{Forward approach}
\label{App:AIC}

The forward approach can be described in the following steps:
\begin{enumerate}
\item Variable selection:
    \begin{enumerate}
 \item Fit $p$ simple linear models: $\overline{y}_c^{(s)} \sim \mathcal{N}\left(\eta_0^{(0)} + \overline{x}^{(s)}_{cj}\eta_1^{(0)}, \sigma^{(0)^2}/n_c\right), \ c=1, \dots, m, \ j=1, \dots, p,$ and compute the $p$ corresponding AICs;
        \item Select $\overline{x}_{k}$ which minimises the AIC;
        \item Fit $p-1$ linear models: $\overline{y}_c^{(s)} \sim \mathcal{N}\left(\overline{\bm{z}}_{(1)c}^{(s)^\top} \bm{\eta}^{(1)}, \sigma^{(1)^2}/n_c\right), \ \overline{\bm{z}}_{(1)}^{(s)} = \left[1, \overline{x}_{k}^{(s)}, \overline{x}_{j}^{(s)}\right]^\top,  \ c=1, \dots, m, \ j=1, \dots, p, \ j \neq k,$ and compute the corresponding AIC;
        \item Select $\overline{x}_{k'}$ which minimises the AIC;
        \item Repeat steps (c) and (d) until the AIC is no longer minimised or while the number of selected variables, $K < m$.
    \end{enumerate}
    \item Final model and prediction:
    \begin{enumerate}
        \item Fit $\overline{y}_c^{(s)} \sim \mathcal{N}\left(\overline{\bm{z}}_c^{(s)^\top}\bm{\eta}, \sigma^2/n_c\right), \ c=1, \dots, m$ to estimate $\widehat{\bm{\eta}}$, $\widehat{\mathbb{V}}\left(\widehat{\bm{\eta}}\right)$ and $\widehat{\sigma}$;
        \item Predict $\widehat{\overline{Y}}_c^{(ns)} = \overline{\bm{z}}_c^{(ns)^\top}\widehat{\bm{\eta}}, \ c=1, \dots, M$;
        \item Estimate the prediction variance, $\widehat{\mathbb{V}}\left(\widehat{\overline{Y}}_c^{(ns)}\right) = \overline{\bm{z}}_c^{(ns)^\top}\widehat{\mathbb{V}}\left(\widehat{\bm{\eta}}\right)\overline{\bm{z}}_c^{(ns)} + \widehat{\sigma}^2/(N_c-n_c), \ c=1, \dots, M$.
    \end{enumerate}
\end{enumerate}

\section{Random Forest algorithm}
\label{App:RF}

A random forest procedure can be described through the following algorithm:
\begin{enumerate}
    \item Draw a bootstrap dataset $\mathcal{D}^{(b)} = \left\{\left(\overline{y}_c^{(s)(b)}, \overline{\bm{x}}_c^{(s)(b)}\right), \ c=1, \dots, m\right\}$;
    \item Train the $b$th tree $T^{(b)}$ using $\mathcal{D}^{(b)}$ with hyperparameters $mtry$ and $nodesize$:
    \begin{enumerate}
        \item Let all responses gather in a single node, $\mathcal{A}$;
        \item Randomly select $mtry \leq p$ covariates. Partition $\mathcal{A}$ into nodes $\mathcal{A}_1$ and $\mathcal{A}_2$ based on $\overline{x}_{j} \leq c$ and $\overline{x}_{j} > c$, respectively, for $\overline{x}_{j}$ one of the $mtry$ selected covariates. The covariate and splitting rule are chosen such that $$\sum_{a=1}^2\sum_{c \in \mathcal{A}_a} \left(\overline{y}_c^{(s)(b)} - \left((1/|\mathcal{A}_a|)\sum_{c' \in \mathcal{A}_a} \overline{y}_{c'}^{(s)(b)}\right)\right)^2$$ is minimised.
        \item Repeat step (b) until there are a maximum of $nodesize$ responses in each final node. 
        \item For a new data point with covariate vector $\overline{\bm{x}}$, $T^{(b)}$ yields a point estimate equal to the mean responses in the final node that corresponds to $\overline{\bm{x}}$: $\widehat{\overline{Y}}^{(b)}(\overline{\bm{x}})=\sum_{c=1}^m w_c^{(b)}(\overline{\bm{x}})\overline{y}_c^{(s)},$ where $w_c^{(b)}, \ c=1, \dots, m,$ are weights associated to the outcome sampled means based on the $b$th bootstrap dataset and tree.
    \end{enumerate}
    \item Repeat steps 1. and 2. $B$ times.
    \item For a new data point with covariate vector $\overline{\bm{x}}$, the random forest yields a point estimate equal to the average over the $B$ point estimates obtained from the $B$ trees: $\widehat{\overline{Y}} = (1/B)\sum_{b=1}^B \widehat{\overline{Y}}^{(b)}(\overline{\bm{x}}) = \sum_{c=1}^m w_c(\overline{\bm{x}}) \overline{y}_c^{(s)}$, where $w_c = (1/B)\sum_{b=1}^B w_c^{(b)}(\overline{\bm{x}}), \ c=1, \dots, m$.
\end{enumerate}

\section{\texttt{R} code: proposed scaled split conformal procedure}
\label{App:SC_Code}

\begin{lstlisting}[caption=R code to obtain prediction intervals associated with random forest estimates through the proposed scaled split conformal procedure,
  label=code]
  # Sample_data: {(xbar_c^s, ybar_c^s), c=1, ..., m}
  # eas: sampled areas
  # Pop_data: {(xbar_c^ns, ybar_c^s), c=1, ..., M}, with ybar_c^s=0 if c is not sampled
  
  # Step 1: split sample data and organise dataset
  selected_eas_half = sample(eas, length(eas)/2, replace = FALSE)
  
  data_to_train = Sample_data %>% filter(ea %in% selected_eas_half)
  data_to_get_residuals = Sample_data %>% filter((!ea %in% selected_eas_half))
  data_to_get_final_estimates = Pop_data
  
  Full_Data = bind_rows(data_to_train,
                        data_to_get_residuals,
                        data_to_get_final_estimates)
  
  # Step 2: train a RF on S1 and predict on S2
  rf = ranger(y ~ ., data = Full_Data[1:nrow(data_to_train),],
              mtry=2, min.node.size = 5, num.trees = 1000, keep.inbag = TRUE)

 all_pred = predict(rf, data = Full_Data[(nrow(data_to_train)+1):nrow(Full_Data),])

  # Step 3: Compute the scaled absolute residuals
  residuals = Sample_data %>% 
    filter((!ea %in% selected_eas_half)) %>% 
    mutate(pred = (all_pred$predictions)[1:nrow(data_to_get_residuals)],
           R_c_scaled = abs(y - pred)*sqrt(n_c))

  # Step 4: Find d_alpha, the relevant quantile for a (1-alpha)% level prediction interval
  d_95 = sort(residuals$R_c_scaled)[ceiling((length(selected_eas_half) + 1)*(1 - 0.05))]
  d_80 = sort(residuals$R_c_scaled)[ceiling((length(selected_eas_half) + 1)*(1 - 0.2))]
  d_50 = sort(residuals$R_c_scaled)[ceiling((length(selected_eas_half) + 1)*(1 - 0.5))]

  # Step 5: Compute the (1-alpha)% level prediction intervals
  predictions = Pop_Data %>%
    mutate(pred = (all_pred$predictions)[(nrow(data_to_get_residuals)+1):length(Pop_Data)],
            y_bar_hat = f_c*y_bar_s + (1-f_c)*pred,
            
            CI_l_95 = f_c*y_bar_s + (1-f_c)*(pred - d_95/sqrt(N_c-n_c)),
            CI_u_95 = f_c*y_bar_s + (1-f_c)*(pred + d_95/sqrt(N_c-n_c)),
            
            CI_l_80 = f_c*y_bar_s + (1-f_c)*(pred - d_80/sqrt(N_c-n_c)),
            CI_u_80 = f_c*y_bar_s + (1-f_c)*(pred + d_80/sqrt(N_c-n_c)),
            
            CI_l_50 = f_c*y_bar_s + (1-f_c)*(pred - d_50/sqrt(N_c-n_c)),
            CI_u_50 = f_c*y_bar_s + (1-f_c)*(pred + d_50/sqrt(N_c-n_c)))
\end{lstlisting}

\section{Extra simulation scenarios: prediction methods comparison}
\label{App:Sim}

This section is a continuation of the simulation study shown in Section \ref{sec:Sim_comparison}. The same finite populations A, B and C are created and, from each, $R=100$ samples are drawn following the three sampling schemes:
\begin{enumerate}
    \item (Stratified) Select all $m=M=500$ areas (EAs) and within each area, sample $n_c = 0.5N_c, \ c=1, \dots, m$ units;
    \item (One-stage) Sample $m = M/2$ areas and within each area, select all $n_c = N_c, \ c=1, \dots, m$ units;
    \item (Two-stage) Sample $m=M/2$ areas and within each area, sample $n_c = 0.5N_c, \ c=1, \dots, m$ units.
\end{enumerate}

Again, for each scenario, the estimates and their prediction intervals are computed as described in Section \ref{sec:Model}, knowing and anonymising the EAs. 

\begin{figure}[!htb]
    \centering
    \includegraphics[width=\textwidth]{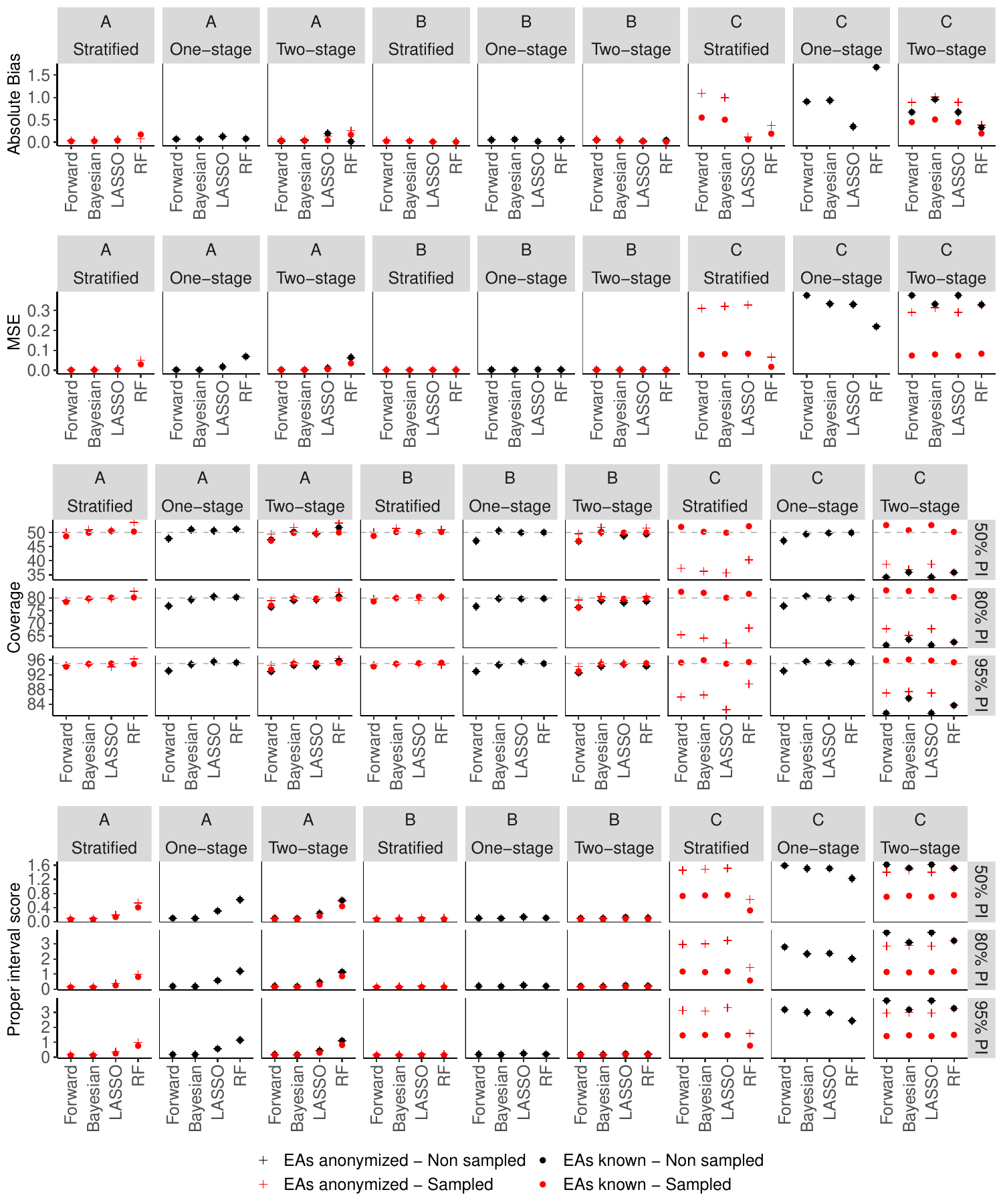}
    \caption{Mean absolute bias, MSE, coverages and proper scores of the prediction intervals, obtained for each method across the 9 simulation scenarios. Forward: forward selection approach; Bayesian: Bayesian shrinkage approach; RF: Random forest approach}
    \label{fig:Sim3_comparison_Results}
\end{figure}

Figure \ref{fig:Sim3_comparison_Results} presents the results (mean absolute bias, MSE, prediction interval coverage and proper score) for each of the 9 simulation scenarios. Like the results in Section \ref{sec:Sim_comparison}, all four modelling methods perform similarly in populations A and B, where the association between the outcome and the covariates is linear. All four methods yield virtually no bias and prediction intervals of the right coverage rate. 

In population C, when the sampled EAs are anonymised, all four methods perform slightly worse than when the sampling information is known: the MSE is multiplied by a factor of 3 and the prediction intervals show under-coverage. When the EAs are anonymised, in all three sampling schemes, the random forest approach leads to smaller MSE and proper interval scores. In terms of MSE, regardless of the sampling design, the random forest performs better than the other three modelling methods, knowing or ignoring which EAs have been sampled. 

\end{appendix}

\end{document}